\documentclass[a4paper, 11pt, oneside]{article} % A4 paper size, default 11pt font size and oneside for equal margins

\usepackage{geometry}
\usepackage{authblk}
\usepackage[utf8]{inputenc} % Required for inputting international characters
\usepackage[T1]{fontenc} % Output font encoding for international characters
\usepackage{fouriernc} % Use the New Century Schoolbook font
\usepackage{epsfig}

\usepackage{cite}

\usepackage{graphicx}
\usepackage{algorithm, algorithmic}
\usepackage{amssymb}
\usepackage{amsmath}
\usepackage{multicol}
\usepackage{float}
\usepackage{subfigure}
\usepackage{threeparttable}
\usepackage{url}
\usepackage{color}
\usepackage{ulem}

\begin{document} 

%%%%封面内容编辑%%%%
\begin{titlepage} % Suppresses headers and footers on the title page

	\centering % Centre everything on the title page
	
	\scshape % Use small caps for all text on the title page
	
	\vspace*{\baselineskip} % White space at the top of the page
	
	%------------------------------------------------
	%	Title
	%------------------------------------------------
	
	\rule{\textwidth}{1.6pt}\vspace*{-\baselineskip}\vspace*{2pt} % Thick horizontal rule
	\rule{\textwidth}{0.4pt} % Thin horizontal rule
	
	\vspace{0.75\baselineskip} % Whitespace above the title
	
	{\LARGE Clustering Residential Electricity Load Curves\\ via \\Community Detection in Network} % Title
	
	\vspace{0.75\baselineskip} % Whitespace below the title
	
	\rule{\textwidth}{0.4pt}\vspace*{-\baselineskip}\vspace{3.2pt} % Thin horizontal rule
	\rule{\textwidth}{1.6pt} % Thick horizontal rule
	
	\vspace{2\baselineskip} % Whitespace after the title block
	
	%------------------------------------------------
	%	Subtitle
	%------------------------------------------------
	
	%Subtitle here % Subtitle or further description
	
	\vspace*{3\baselineskip} % Whitespace under the subtitle
	
	%------------------------------------------------
	%	Editor(s)
	%------------------------------------------------
	
	Edited By
	
	\vspace{0.5\baselineskip} % Whitespace before the editors
	
	{\scshape\Large Yunyou Huang\\ Jianfeng Zhan\\ Nana Wang\\ Chunjie Luo\\ Lei Wang\\ Rui Ren} % Editor list
	
	\vspace{0.5\baselineskip} % Whitespace below the editor list

	\vfill % Whitespace between editor names and publisher logo
	
	%------------------------------------------------
	%	Publisher
	%------------------------------------------------
	
	%\plogo % Publisher logo
	%\def\BUlogo{\epsfig{file=ICT.pdf,height=3cm}}
	%\includegraphics[scale=0.135]{ICT.pdf}
	\epsfig{file=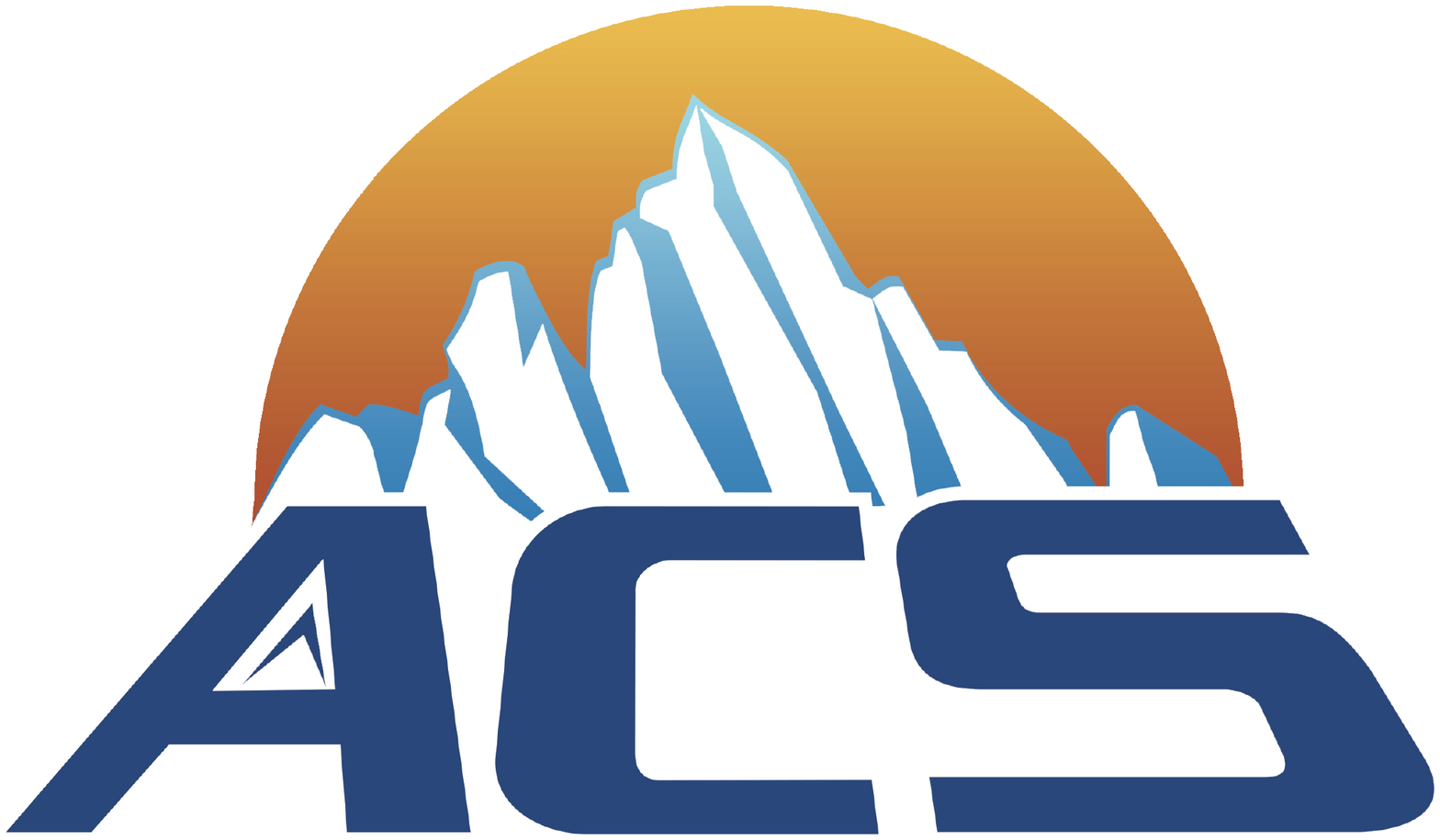,height=5cm,angle=270}
	\textit{\\Software Systems Laboratory (SSL), ACS\\ICT, Chinese Academy of Sciences\\Beijing, China\\http://prof.ict.ac.cn/ssl} % Editor affiliation
	\vspace{5\baselineskip} % Whitespace under the publisher logo

	Technical Report No. ACS/SSL-2018-5 % Publication year
	
	{\large November 23, 2018} % Publisher

\end{titlepage}

%----------------------------------------------------------------------------------------

%%%title here%%%
\title{Clustering Residential Electricity Load Curves via \\Community Detection in Network}

\author{Yunyou Huang, Jianfeng Zhan, Nana Wang, Chunjie Luo, Lei Wang and Rui Ren}

\date{November 27, 2018}
\maketitle

%%%%abstract here
\begin{abstract}
Performing analytic of household load curves (LCs) has significant value in predicting individual electricity consumption patterns, and hence facilitate developing demand-response strategy, and finally achieve energy efficiency improvement and emission reduction. LC clustering is a widely used analytic technology, which discovers electricity consumption patterns by grouping similar LCs into same sub-groups. However, previous clustering methods treat each LC in the data set as an individual time series, ignoring the inherent relationship among different  LCs, restraining the performance of the clustering methods. What's more, due to the significant volatility and uncertainty of LCs, the previous  LC clustering approaches inevitably result in either lager number of clusters or huge variances within a cluster, which is unacceptable for actual application needs. In this paper, we proposed an integrated approach to address this issue. First, we converted the LC clustering task into the community detection task in network. Second, we proposed a clustering approach incorporated with community detection to improve the performance of LC clustering, which includes network construction, community detection and typical load profile  extraction. The method convert the data set into a network in which the inherent relationship among LCs is represented by the edges of the network. Third, we construct a multi-layer typical load profile directory to make the trade-off between variances within a cluster and the number of the clusters, enabling the researchers to assess the LCs and the customers in different layers according to practical application requirements. The experiments demonstrate that our integrated approach outperform the state-of-the-art methods.

\end{abstract}

\clearpage

\section{Introduction}
\begin{figure*}
\centering
\includegraphics[width=5in, height=1.5in]{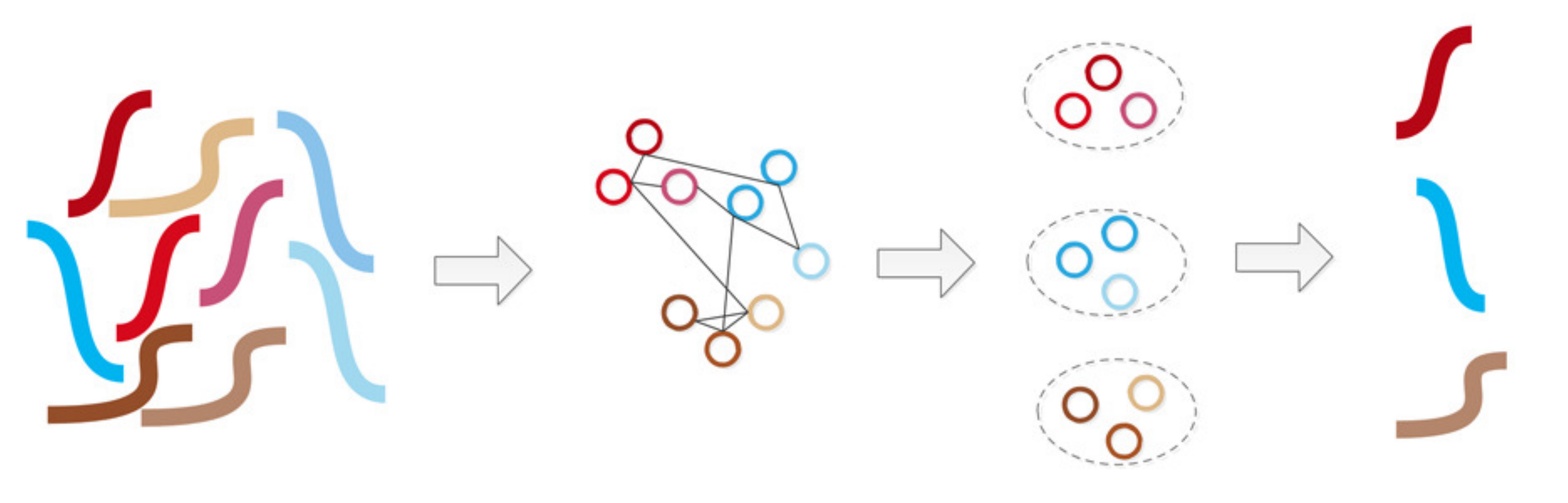}
\caption{\textbf{The summary of our clustering approach incorporated with community detection.} LCs clustering via community detection in networks.~\label{Fig1}}
\end{figure*}
    With the rapid development of economy and society, the electricity consumption has increased to 5802.00 billion kWh in 2015 in China, the 74\% of which is provided by the thermal power plant~\cite{zhou2016understanding, NationalData}. A large amount of greenhouse gases are emitted by the thermal power plant, aggravating the environmental pollution. It is becoming more pressing to increase energy efficiency and reduce emission. Analyzing the electricity consumption of the customers is an essential step in the development of demand-response strategies to curtail electricity demand and keep balance between supply and demand by load shifting, peak clipping, load profile reshaping~\cite{alasseri2017review, lund2003management}, and achieve energy efficiency improvement and emission reduction.
    The electricity consumption of the customers is recorded by the smart meter as  household LCs and provides rich information of the electricity consumption behaviors and lifestyles of the customers~\cite{wang2018review}. The knowledge, such as typical load profiles ( in short, TLP) extracted from household LCs,  achieves  a better understanding of occupancy behavior to infer habitual consumption patterns and identify potential energy efficiency and demand-response options, which in turn help network operators with demand-side management strategies to increase energy efficiency and reduce emissions~\cite{yildiz2017recent}.

    The clustering methods, the most common technology of LC analyzing, is an unsupervised method to discover the unknown knowledge (such as the TLPs) in the data set by grouping the similar LCs into same sub-groups. Various clustering methods, such as K-means, fuzzy clustering, hierarchical clustering, and self-organizing maps, have been applied to LCs clustering~\cite{chicco2006comparisons, yang2013review, yildiz2017recent, wei2018review, wang2018review}. To improve the performance of LC clustering, in the recent state-of-the-art work,  the Dynamic Time Warping (DTW) is recently introduced to measure the distances between the LCs, and combine the K-medoids to cluster the LCs~\cite{teeraratkul2017shape}. Unfortunately, the previous LC clustering methods  still have the following weakness.
     % However, the performance of clustering methods are commonly restrained by not only the distance measure but also the clustering algorithms~\cite{granell2015impacts, lin2017clustering}.
    %There is still a lot room for improvement. In addition, the clustering methods inevitably results in either lager number of cluster or huge variances within cluster, due to the significant volatility and uncertainty of the LCs~\cite{li2016multi,teeraratkul2017shape}, hardly directly meeting the actual application needs that requires the trade-off between number of clusters and variances within clusters.
     %The previous LC clustering methods applied to the LCs clustering have the following challenges.
     %When the clustering methods are used to cluster the LCs, only the local relationship among the neighbor LCs are able to be easily identified, and the long distance global relationship remains unknown in general\cite{ferreira2016time}. In other words, the direct relationship between LCs is easily identified, and the indirect relationship between the LCs and the relationship between the groups of LCs are hardly to be identified. Thus, a powerful method is required to synchronously handle both the local and global relationship between the LCs.
\begin{itemize}
\item[a)] It restrains the performance of the clustering methods in the way that each LC in the data set is treated as an individual time series and hence the inherent relationships among  LCs are ignored.
\item[b)] Though the state-of-the-art clustering method that adopts the DTW~\cite{teeraratkul2017shape} performs better than the LC clustering approach that used Euclidean distance as the distance measure, it is unable to obtain the cluster centers that represents the TLP in the common averaging method, and hence hampers the consumer electricity consumption pattern extraction.
\item[c)] Due to the significant volatility and uncertainty of the LCs~\cite{li2016multi,teeraratkul2017shape},  the clustering methods inevitably results in either a lager number of clusters or huge variances within a cluster, which is unacceptable for the actual application needs that require the tradeoff between the number of clusters and variances within a cluster.
\end{itemize}

    %community detection incorporated
    %Thus, the LCs clustering is an challenging task.
    In this paper, we proposed an integrated approach to address the above issue. Our approach includes two integrated parts: a new \underline{c}lustering method \underline{i}ncorporated with \underline{c}ommunity \underline{d}etection (in short, CICD) to improve the LC clustering performance, and a  best-cluster-number-determined approach to make the trade-off between variances within a cluster and the number of the clusters.

    As shown in Figure~\ref{Fig1}, CICD consists of network construction, community detection, and typical load profile (TLP) extraction.  First, we converts the LCs data set into a $\varepsilon-$nearest neighbor network ($\varepsilon\verb|-|\!N\!N$) using the distance measure DTW, characterizing both local and global inherent relationship between any pair or any groups of LCs. Second, we employ a  modularity-based algorithm Louvain to synchronously optimize the local and global modularity, and obtain the optimal community partition, where a community represents a cluster. Third, we extract the centers, each of which represents a TLP, from clusters using the averaging method---DTW Barycenter Averaging---to obtain the typical electricity consumption patterns of the customers. Compared with the K-medoids\&DTW clustering method in the~\cite{teeraratkul2017shape}, our method has significant improvement, using the metric of the common cluster validity indices.

     For the best-cluster-number-determined approach, we firstly segments the cluster number into $n$ intervals. Secondly, for each interval $[i, j)$, we select a value $k (i\leq k < j)$ as the best cluster number based on the performance of CICD with the cluster number $k$. Thus, we are able to construct a multi-layer TLP directory, a layer of which is a clustering result of CICD with the best cluster number $k$ in an interval. In each layer of the TLP directory, the variance within the cluster is small when the number of the the TLPs is large, and the variance within the cluster is large when the number of the the TLPs is small. In all the layers, the variance within the cluster decrease when the number of the TLPs increases. Thus, the TLP directory has the ability to provide the trade-off between variances within a cluster and the number of the cluster, and enables the researcher to assess the LC and the customer in different layers according to practical application requirements.

     %The TLP directory provides the trade-off between variances within cluster and the number of the cluster, and enable the researcher to assess the LC or the customer in different layers according to practical application requirements.

    The paper is structured as follows. Section 2 summarizes the previous LC clustering methods. Section 3 introduces our innovative approach. Section 4 presents and discusses the results. Section 5 introduces the TLP construction method. Section 6 draws a concluding remark.

\section{Related Work}
Various methods are applied to the LCs clustering, such as K-means, self-organizing maps and hierarchical clustering. These common used methods are roughly grouped into four categories according to the clustering criterion: partitioning method, hierarchical method, density-based method and model-based method~\cite{rasanen2010data,chicco2012overview, yang2013review, ramos2015data, costacurta2017application,yildiz2017recent, rajabi2017review, wang2018review, wei2018review}.

    The partitioning methods is the most common used clustering technology in the LCs clustering, due to the simplicity and low time complexity. The partitioning methods initially select $K$ centroids, and iteratively update these points to optimize the cost function~\cite{ramos2015data}. These methods include K-means~\cite{kwac2014household, rhodes2014clustering,quilumba2015using, lavin2015clustering, zhang2016cluster,du2016customer,al2017k}, K-medoids~\cite{teeraratkul2017shape}, and fuzzy C-means~\cite{yang2010study,nikolaou2012application,selvam2017fuzzy}.

    The hierarchical methods applied to the LCs clustering includes two types: agglomerative and division, the agglomerative of which is the most commonly used technology~\cite{wei2018review}. For the agglomerative methods, each LC is firstly initialized to a cluster. Then, the two closest clusters are combined into a new cluster, thus reducing the number of clusters in~\cite{jota2011building}. In addition,~\cite{kwac2014household} employed the hierarchical clustering to merge the clusters generated by the K-means, and reduce the number of the clusters.

    The density-based method consider the LC in high-density regions in space as clusters, and the ones in low-density regions as outliers or noise~\cite{ramos2015data}. In~\cite{wang2016clustering},  a density-based method, which cluster the objects by fast search and find of density peaks algorithm~\cite{rodriguez2014clustering}, cluster the customers, using the daily load curves. In~\cite{jin2017comparison}, the classic density-based method Density-Based Spatial Clustering of Applications with Nois is applied to cluster daily residential meter data.

    The model-based methods assume a model for each clusters, and find the best fit of the data to the given model~\cite{ramos2015data}. In~\cite{labeeuw2013residential,haben2016analysis,li2016multi}, the finite mixture models Gaussian mixture models are applied to cluster the smart meter data. In~\cite{chicco2003application,mcloughlin2015clustering}, the self-organizing maps method clusters the smart meter data.

     In addition to the method above, some new methods, such as spectral clustering~\cite{sanchez2014hierarchical, lin2017clustering}, hierarchical K-means~\cite{xu2017hierarchical}, Support Vector Clustering~\cite{gavrilas2010application}, and the iterative self-organizing data-analysis technique algorithm~\cite{mutanen2011customer} have also been applied to the LCs clustering.

     For the LCs clustering method above, each LC in the data set is treated as an individual time series, ignoring the inherent relationship between LCs, which restrain the performance of the clustering methods. The network is a powerful mechanism to characterize the time series data set that has been proved on the time series analysis~\cite{ferreira2016time}. This paper fills the research gap mentioned above by converting the LCs data set into a network and clustering the LCs via the community detection in network, achieving the improvement over the state-of-the-art method.

\section{Methodology}

     In this section, we give a detailed description of our method, which consists of 4 steps: data preparation, network construction, community detection, center extraction.

\subsection{Data preparation}

    The normalization is an indispensable step in the LCs clustering, as it shields the amplitude interference and make the user consumption pattern contained in the LCs to be easily identified. In our work, for a given daily LC $s_t$, we obtained the normalized daily LC $l_t$ by the Equation~\ref{eq1}.

\begin{equation}\label{eq1}
  l_t=\frac{s_t}{a}   \quad and  \quad a=\sum_{t=1}^{96} s_t
\end{equation}

\subsection{Network Construction}
\subsubsection{Dynamic Time Warping (DTW)}

    The distance between LCs determine the formation of edges in the network. A most famous shape-based distance measure of the time series DTW is applied to measure the distance between LCs, since the demand response is focused on understanding the electricity consumption patters of the customer. The purpose of the DTW is finding the optimal comparison path between the time series and calculating the distance between the time series~\cite{berndt1994using}. For two given LCs $X=\{x_0, x_1, ... , x_{n-1}\}$ and $Y=\{y_0, y_1, ... , y_{n-1}\}$, the warping path $P=\{p_0, p_1, ... , p_k\}$, $p_k=(i_k,j_k) \in [0 : n-1] \times [0 : n-1]$ is obtained according to the following restrictions~\cite{teeraratkul2017shape}.
\begin{itemize}
\item[a] Boundary condition : $p_0 = (0, 0)$ and $p_k = (n-1, n-1)$.
\item[b] Monotonicity condition : $i_{k-1} \leq i_k$ and $j_{k-1} \leq j_k$.
\item[c] Continuity condition : $i_k - i_{k-1} \leq 1$ and $j_k - j_{k-1} \leq 1$.
\item[d] Warping window : $|i_k -j_k | < w$.
\end{itemize}
     Here, we assume that a typical consumer exhibits less than $1$ hour variation in time of electricity consumption. Customer electricity consumption data are collected every 15 minutes. Thus the warping window $w$ of the DTW  is set to $4$.

    The minimum total cost of the warping path $ c_{P(X,Y)}$ is equal to the distance between the LCs X and Y. And the $c_p$ is obtained by the Equation~\ref{eq2}.
\begin{equation}\label{eq2}
\left.
  \begin{array}{c}
    c_{P(X,Y)}=\eta(n-1,n-1)\\
    \\
    \eta(i,j)=\delta(i,j)\\+min[\eta(i-1,j), \eta(i-1,j-1), \eta(i,j-1)]\\
    \\
     \delta(i,j)=\left\{ \begin{array}{lc}d(x_i,y_j)&|i-j| < 4 \\ Max \quad value& else \end{array}\right.
  \end{array}
\right.
\end{equation}

\subsubsection{The conversion of the LC data set into network}
    The network is a powerful mechanism, and is able to represent the complex relationship between the LCs. A network $G(V,E)$ consist of  $n$ vertices $V=\{v_0, v_1, ... , v_{n-1}\}$ and $m$ edges $E=\{e(v_i,v_j) | v_i, v_j \in V\}$, where the $e=(v_i,v_j)$ is an edge that connects $v_i$ and $v_j$. In this work, we converted the LCs data set into a $\varepsilon-$nearest neighbor network ($\varepsilon\verb|-|\!N\!N$). Firstly, we assigned each LC in the original data set as a vertex. Secondly, the edge between two vectors was formed when the distance between them less than $\varepsilon$ in the Equation~\ref{eq3}.

\begin{equation}\label{eq3}\centering
\left.
  \begin{array}{c}
\varepsilon=\lambda \ast \mu_d \\
\\
 \mu_d=\frac{1}{n}\sum\limits_{i=0}^{n-1}d(x,i)\\
  \end{array}
\right.
\end{equation}
     Here, the $ \mu_d$ is the mean distance of the vector $x$ and all of the other vectors, the $\lambda$ is a parameter to adjust the number of the edge in the network. The higher of the $\lambda$ is, the more edges the network contained and the more complex of the network is. In addition, the weight of the edge $e\in E$ is determined by the Equation~\ref{eq4}.
\begin{equation}\label{eq4}\centering
%\scriptsize
\left.
  \begin{array}{c}
  w_e=1-\frac{d(v_i,v_j)}{d_{max}} \quad and \quad e=(v_i,v_j)\in E\\
  \\
  d_{max}=Max\left[ \;d(v_k,v_l) \;| \;e=(v_k,v_l)\in E  \;\right ]
  \end{array}
\right.
\end{equation}

\subsection{Community Detection}
    The community detection is a very important method that boost the finding of a-priori unknown modules in the network, hence attracting a lot of attention~\cite{blondel2008fast}. In this work, a famous modularity-based algorithm Louvain is applied to extract the communities from the network, as it has the advantage in the computation time and the quality of the communities detection with the metric of the modularity in Equation~\ref{eq5} ~\cite{newman2004analysis,newman2006modularity,blondel2008fast}. As shown in Algorithm~\ref{alg1}, the Louvain iteratively extracted the communities from the network to find the optimal division.
\begin{equation}\label{eq5}\centering
\left.
  \begin{array}{c}
  Q=\frac{1}{2m}\sum\limits_{i,j}[A_{i,j}-\frac{k_ik_j}{2m}]\delta(c_i,c_j)\\
  \\
  m=\frac{1}{2}\sum\limits_{u,v}A_{u,v}\\
  \\
  \delta(c_i,c_j)=\left\{ \begin{array}{lc}1&c_i=c_j \\ 0& else \end{array}\right.
  \end{array}
\right.
\end{equation}
    Here, $A_{i,j}$ is the weight of the edge between vertex $i$ and $j$,  $k_i=\sum A_{i,l}$ is the sum of the weights of the edges attached to $i$, and $c_i$ is the community which $i$ is assigned to.

\begin{algorithm}[!htbp]
         \renewcommand{\algorithmicrequire}{\textbf{Input:}}
         \renewcommand{\algorithmicensure}{\textbf{Output:}}
         \caption{Louvain}
         \label{alg1}
         \begin{algorithmic}[1]
                   \REQUIRE Network G(V,E).
                   \ENSURE Communities $C_{final}$.
                   \STATE Set $Change=True$
                   \WHILE {$Change$}
                   \STATE Set each vertex $v_i\in V$ as a different community $c_i\in C$.
                   \STATE Set $Change=False$
                   \STATE Set $LocalChange=True$
                   \WHILE{$LocalChange$}
                   \STATE Set $LocalChange=Flase$
                   \FORALL{ Vertex $v_j \in V $}
                   \STATE Set the vertices which directly connect to vertex $v_j$ as $V_{nei}$.
                   \FORALL{ Vertex $v_l \in V_{nei}$ }
                   \STATE Set the community of $v_j$ as $c_u$ and the community of $v_l$ as $c_v$.
                   \IF{$c_u \neq c_v$}
                   \STATE Calculating the $\Delta Q$ for $v_j$ by Equation~\ref{eq6}.
                   \ENDIF
                   \ENDFOR
                   \IF{The max value of $\Delta Q$ large than 0}
                   \STATE Move $v_j$ to the $c_v$ which obtain the max $\Delta Q$.
                   \STATE Set $LocalChange=True$
                   \ENDIF
                   \ENDFOR
                    \STATE Set $Change=LocalChange \quad || \quad Change$
                   \ENDWHILE
                   \IF{Change}
                   \FOR{$c_i \in C$}
                   \STATE Merge the $c_i$.
                   \ENDFOR
                   \STATE Construct the new Network G.
                   \ENDIF
                   \ENDWHILE
                   \STATE \textbf{return} $C$
         \end{algorithmic}
\end{algorithm}

\begin{algorithm}[!htbp]
         \renewcommand{\algorithmicrequire}{\textbf{Input:}}
         \renewcommand{\algorithmicensure}{\textbf{Output:}}
         \caption{DTW Barycenter Averaging}
         \label{alg2}
         \begin{algorithmic}[1]
                   \REQUIRE All of the LCs $lc\in C$ and the initial center $\overline{c}$.
                   \ENSURE $\overline{c}$.
                   \STATE Set $CT$ be a table of size $n$ containing in each cell a set of coordinates associated to each coordinate of $\overline{c}$.
                   \WHILE{$\overline{c}$ is not stable}
                   \STATE $CT=[\varnothing, \varnothing, ..., \varnothing]$
                   \FOR{$lc_i\in C$}
                   \STATE Obtain the warping path P=DTW($\overline{c}$,$lc_i$)
                   \FOR{$p_k=(i_k,j_k)\in P$}
                   \STATE Put $lc_i[j_k]$ to $CT[i_k]$.
                   \ENDFOR
                   \ENDFOR
                   \FOR{$0\to n-1$}
                   \STATE $ct=CT[i]$
                   \STATE c[i]=$\left.\frac{\sum\limits_{x\in ct}x}{|ct|} \right.$
                   \ENDFOR
                   \ENDWHILE
         \end{algorithmic}
\end{algorithm}

\begin{equation}\label{eq6}\centering
\left.
  \begin{array}{c}
  \Delta Q=
  [\frac{\sum_{in}+\gamma k_{j,in}}{2m}-(\frac{\sum_{tot}+k_j}{2m})^2]\\
  \\
 \quad \quad \quad    -[\frac{\sum_{in}}{2m}-(\frac{\sum_{tot}}{2m})^2-(\frac{k_j}{2m})^2]
  \end{array}
\right.
\end{equation}

    Here, the $\sum_{in}$ is the sum of the weights of the edges inside community $c_v$, $\sum_{tot}$ is the sum of the weights of the edges incident to the vertex inside $c_v$, $k_i$ is the sum of the weights of the edges connect to vertex $i$, $k_{i,in}$ is the sum of the weights of the edges from $i$ to the vertex in $c_v$, and the $m$ is the sum of weights of the network~\cite{blondel2008fast}. The $\gamma$  is a parameter that adjust the number of the communities. Commonly, the smaller the $\gamma$  is, the more communities is extracted from the network.

\subsection{Center Extraction}

    The center of a LC cluster is a TLP which reflects the electricity consumption pattern, and it usually obtained by an averaging method. However, when the DTW is used as the distance measure, the LCs averaging is a difficult task, because it has to be consistent with the ability of DTW to realign sequences over time~\cite{petitjean2011global}. In this paper, a averaging technology DTW Barycenter Averaging shown in Algorithm~\ref{alg2} is proposed by~\cite{petitjean2011global}, and is applied to extract the TLPs from the clusters.
\section{Results and discussion}
    To verify the efficiency of the method, the clustering method are validated on the same data set in the same task.

\subsection{The cluster validity indices}

     It is significant for the clustering algorithm to select the indexes which reflect the performance of algorithms. %There are 30 common cluster validity indices(CVIs) studied in ~\cite{arbelaitz2013extensive}. Inspired
     According to the performance of each common cluster validity index on various data sets~\cite{arbelaitz2013extensive, zhou2018novel}, we selected the 5 cluster validity indices to evaluate the result of clustering: Davies-Bouldin index (DB)~\cite{davies1979cluster}, VCN index (VCN, an improvement of Silhouette)~\cite{zhou2018novel}, S\_Dbw Index (S\_Dbw)~\cite{halkidi2001clustering}, Score function (SF)~\cite{saitta2007bounded}, and COP index (COP)~\cite{gurrutxaga2010sep}. In addition, the $entropy$, a common indexes to identify the demand response potential customers, is also considered to measure the consumer variability in our work~\cite{teeraratkul2017shape}. The smaller $DB$, $S\_Dbw$ and $COP$ are, the higher $VCN$ and $SF$ are, the higher cluster quality is.

DB
\begin{equation}\label{eq7}\centering
\left.
  \begin{array}{c}
  DB=\frac{1}{k}\sum\limits_{c_i\in C}\max\limits_{c_j\in C, c_j\neq c_i}\{\frac{S_{c_i}+S_{c_j}}{d(\overline{c_i},\overline{c_j})}\}\\
  \\
  S_{c_i}=\frac{1}{|c_i|}\sum\limits_{x_i\in c_i}d(x_i,\overline{c_i})
  \end{array}
\right.
\end{equation}

VCN
\begin{equation}\label{eq8}\centering
\left.
  \begin{array}{c}
  VCN=\frac{1}{k}\sum\limits_{i=1}^{k}\frac{bd(c_i)-wd(c_i)}{\max\{bd(c_i),wd(c_i)\}}\\
  \\
  bd(c_i)=\min\limits_{1\leq j\leq k}(\frac{1}{|c_i|}\sum\limits_{l=1}^{|c_i|}d(x_l,\overline{c_j}))\\
  \\
  wd(c_i)=\frac{1}{|c_i|}\sum\limits_{l=1}^{|c_i|}d(x_l,\overline{c_i})
  \end{array}
\right.
\end{equation}

SF
\begin{equation}\label{eq10}\centering
\left.
  \begin{array}{c}
  SF=1-\frac{1}{e^{e^{bcd+wcd}}}\\
  \\
  bcd=\frac{\sum\limits_{i=1}^kd(\overline{c_i},\overline D)*|c_i|}{n*k}\\
  \\
  wcd=\sum\limits_{i=1}^k\frac{1}{|c_i|}\sum\limits_{x\in c_i}d(x,\overline{c_i})
  \end{array}
\right.
\end{equation}

COP
\begin{equation}\label{eq11}\centering
\left.
  \begin{array}{c}
  COP=\frac{1}{n}\sum\limits_{c_i\in C}|c_i|\frac{\frac{1}{|c_i|}\sum\limits_{x_l\in c_i}d(x_l,\overline{c_i})}{\min\limits_{x_m\notin c_i}\max\limits_{x_l\in c_i}d(x_m,x_l)}
  \end{array}
\right.
\end{equation}

Entropy
\begin{equation}\label{eq12}\centering
\left.
  \begin{array}{c}
  S=-\sum\limits_{i=1}^kp_ilog(p_i)
  \end{array}
\right.
\end{equation}
S\_Dbw
\begin{equation}\label{eq9}\centering
\left.
  \begin{array}{c}
  S\_Dbw=Scat+Dens\_bw\\
  \\
  Scat=\frac{\frac{1}{k}\sum\limits_{i=1}^k\sigma_{c_i}}{\sigma_{D}}\\
    \\
  Dens\_bw=\frac{1}{k*(k-1)} \\
  *\sum\limits_{i=1}^k(\sum\limits_{j=1,i\neq j}^k\frac{density(\mu_{i,j})}{\max\{density(\upsilon_i),density(\upsilon_j)\}})\\
  \\
  \sigma_{D}=\frac{1}{|D|}\sum\limits_{x_l \in D}d(x_l,\overline{D})^2\\
  \\
  density(\mu)=\sum\limits_{x_l\in c_i \cup c_j}f(x_l,\mu)\\
  \\
  density(\upsilon_i)=\sum\limits_{x_l\in c_i }f(x_l,\upsilon_i)\\
  \\
  f(x_l,\varphi)=\left\{ \begin{array}{lc}1& d(x,\varphi)>stdev \\ 0 & else \end{array}\right.\\
  \\
  stdev=\frac{1}{k}\sqrt{\sum\limits_{c_i\in C}\sigma_{c_i}}
  \end{array}
\right.
\end{equation}

\subsection{Data Description}
    The data that validate the clustering algorithm is provided by the Pecan Street Inc.~\cite{DataPort}. It contains 22113 daily LCs from 351 households. The LCs are collected from the houses for the period of 63 days between July 6 2015 to September 6 2015, each of which is collected every 15 minutes. In order to obtain the baseline, the data set is also clustered by the state-of-the-art work (K-mediods \& DTW) which is proposed in~\cite{teeraratkul2017shape}.

\subsection{Cluster Performance Evaluation}
\begin{figure}
\centering
\includegraphics[width=2.45in, height=1.4in]{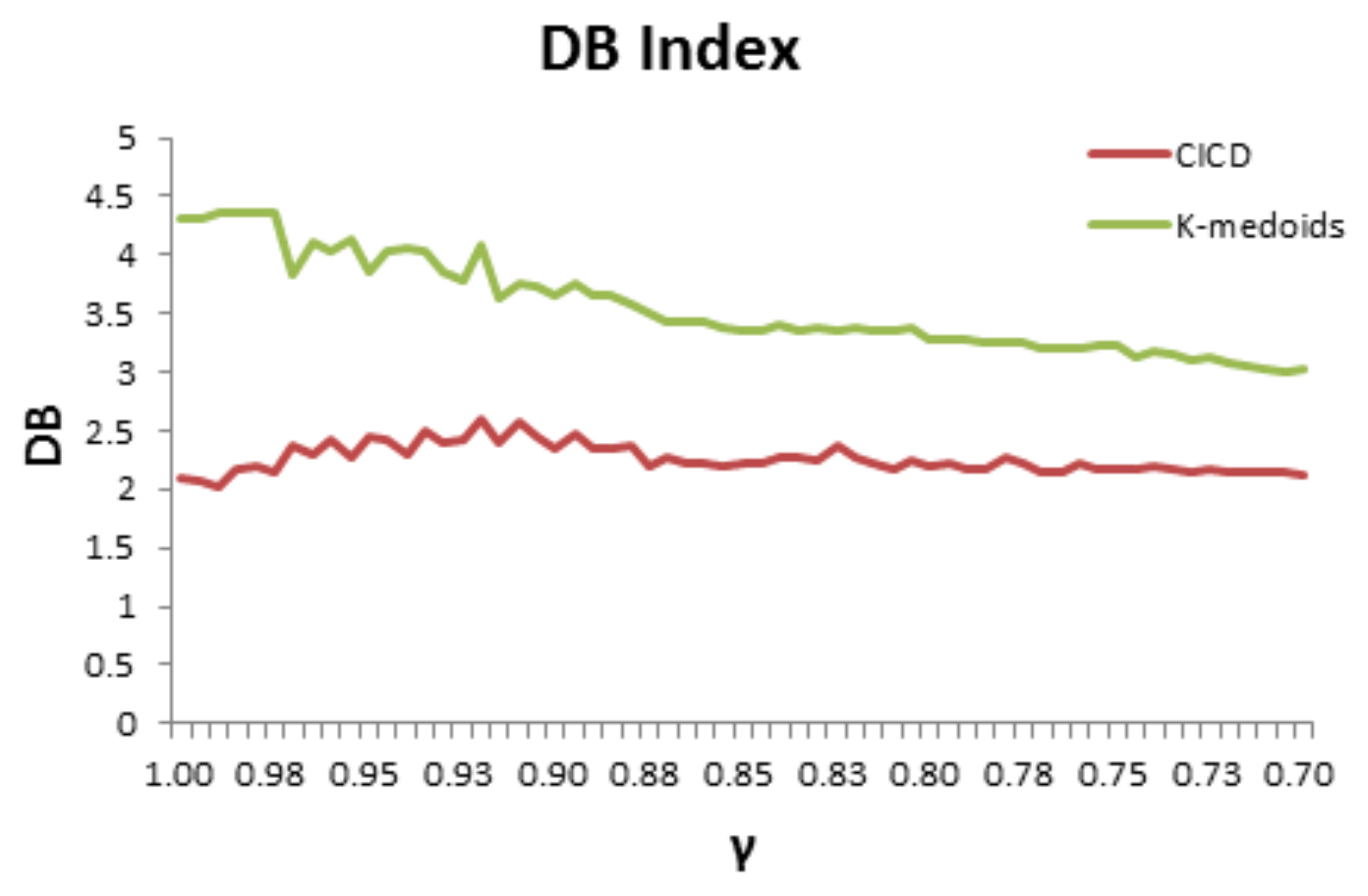}
\caption{\textbf{DB Index.} The CICD is better than K-medoids in DB, especially when the $\gamma$ is large.~\label{Fig2}}
\end{figure}

\begin{figure}
\centering
\includegraphics[width=2.45in, height=1.4in]{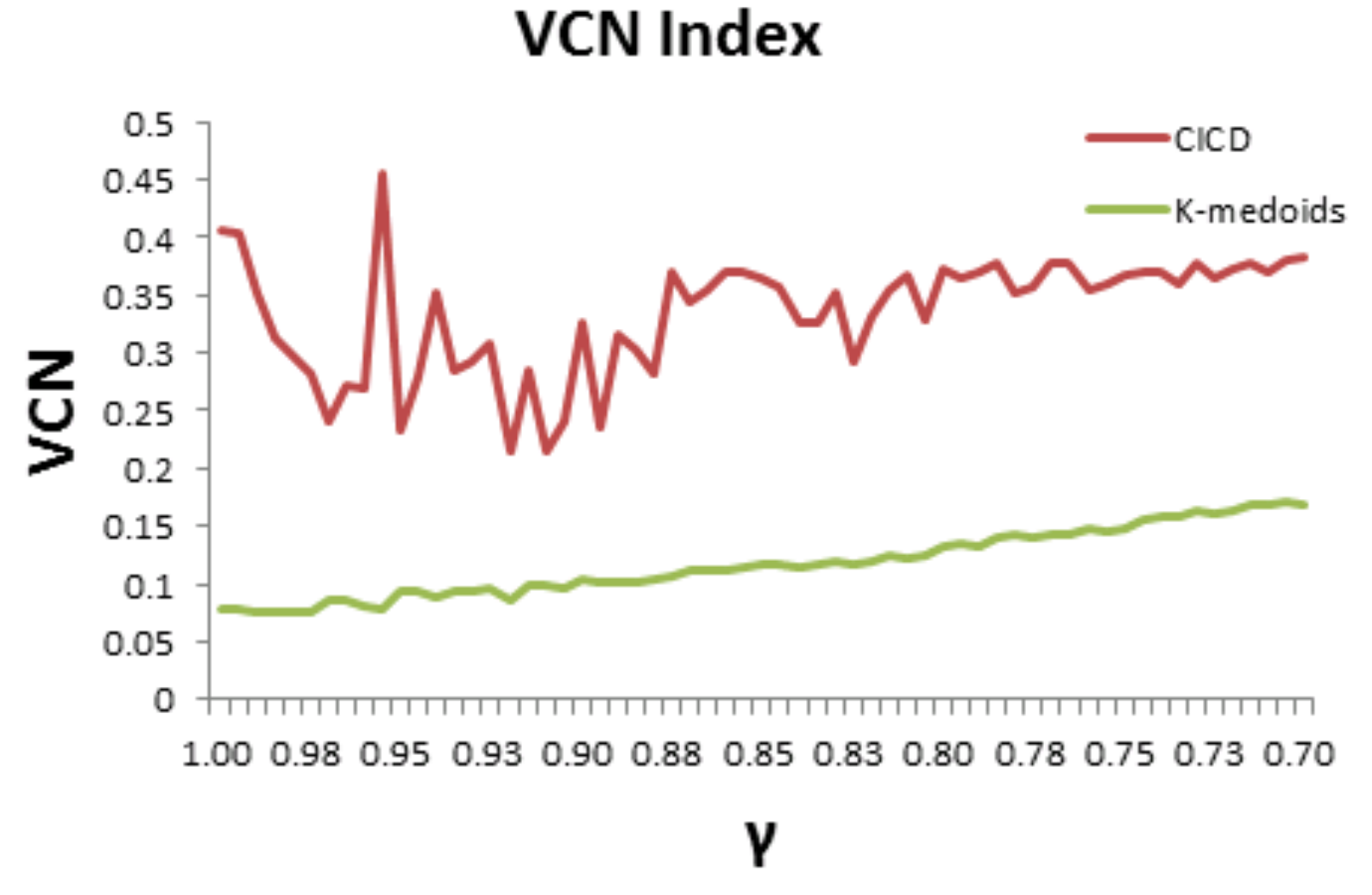}
\caption{\textbf{VCN Index.} The CICD is significant better than the K-medoids in VCN.~\label{Fig3}}
\end{figure}

\begin{figure}
\centering
\includegraphics[width=2.45in, height=1.4in]{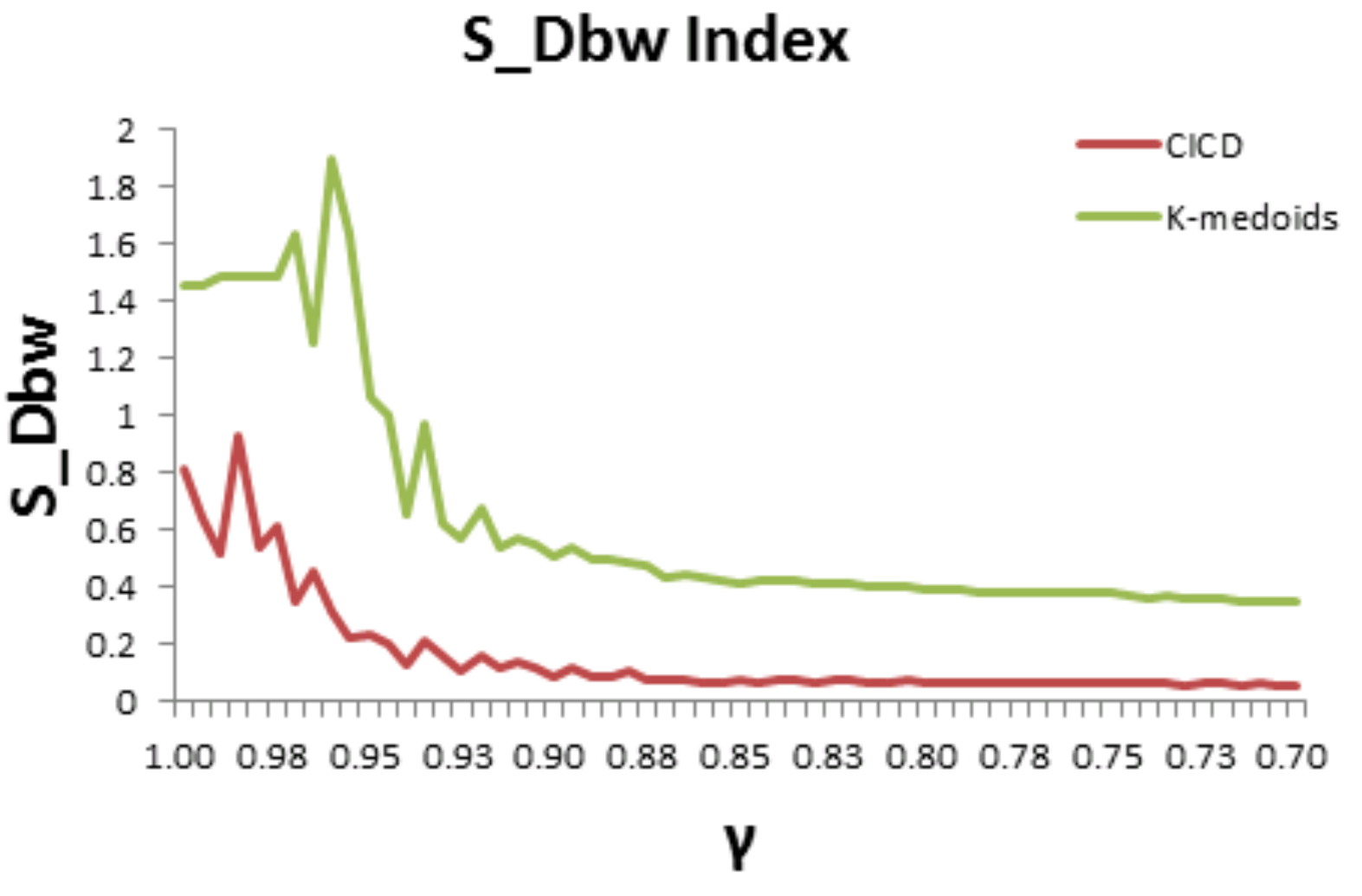}
\caption{\textbf{S\_Dbw Index.} The CICD is better than K-medoids in S\_Dbw, especially when the $\gamma$ is large.~\label{Fig4}}
\end{figure}

\begin{figure}
\centering
\includegraphics[width=2.45in, height=1.4in]{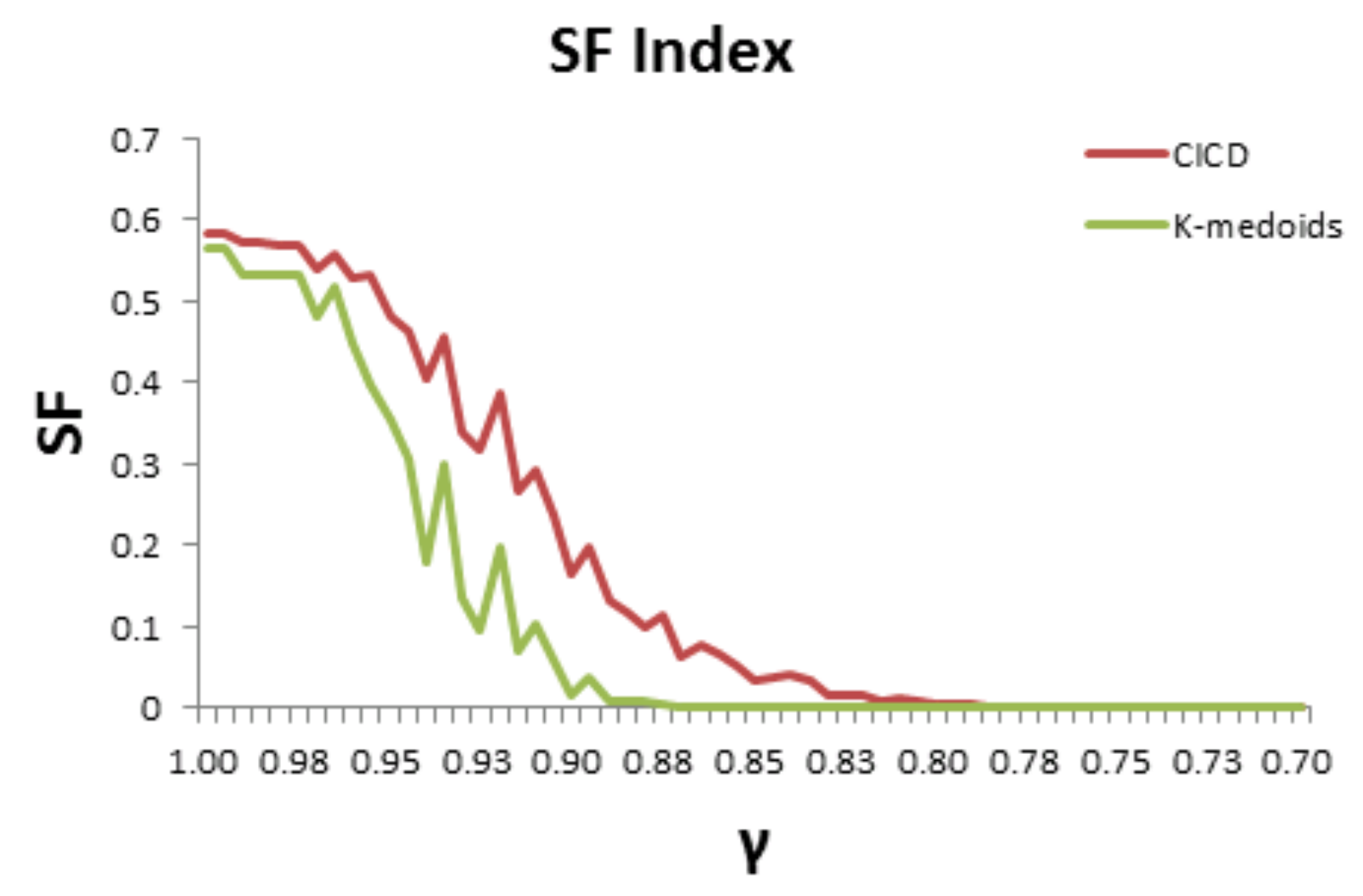}
\caption{\textbf{SF Index.} The CICD is better than K-medoids in SF in most of time, especially when the $\gamma$ is large.~\label{Fig5}}
\end{figure}

\begin{figure}
\centering
\includegraphics[width=2.45in, height=1.4in]{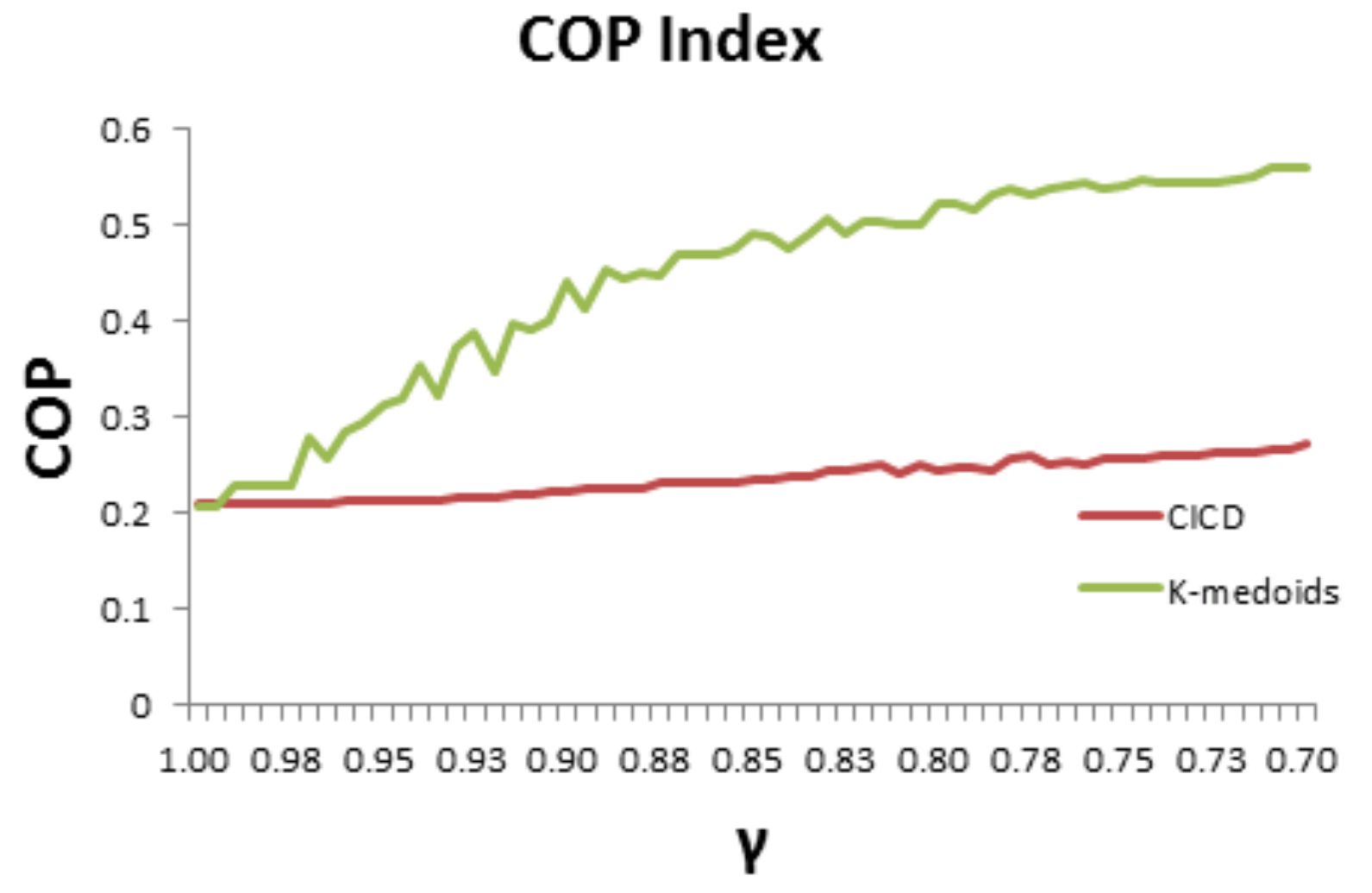}
\caption{\textbf{COP Index.} The CICD is better than K-medoids in COP, especially when the $\gamma$ is small.~\label{Fig6}}
\end{figure}

\newcommand{\tabincell}[2]{\begin{tabular}{@{}#1@{}}#2\end{tabular}}  %è¡šæ Œèªåšæ¢è¡
\begin{table*}[t]
\scriptsize
\centering
\begin{threeparttable}
\caption{The statistics of the cluster validity indices of the clustering results.\label{Tab1}}
\begin{tabular}{ccccccc}
\hline
&$DB$&$VCN$&$S\_Dbw$&$SF$&$COP$&$entropy$\\
\\
$Mean_{K-medoids}$&3.5513&0.1178&0.6397&0.1144&0.4400&2.6156\\
\\
$Mean_{CICD}$&2.2569&0.3347&0.1620&0.1738&0.2350&1.0235\\
\\
Improment&36.45\%&184.13\%&74.68\%&51.92\%&46.59\%&60.87\%\\
\\
\\
$Max_{K-medoids}$&4.3581&0.0788&1.6256&0.1790&0.5289&2.7225\\
\\
$Max_{CICD}$&2.0125&0.4545&0.2245&0.4054&0.2428&0.8476\\
\\
Improment&53.82\%&476.78\%&86.19\%&126.48\%&54.09\%&68.87\%\\

\hline
\end{tabular}
\begin{tablenotes}
\item[*] The $Mean_{K-medoids}$ is the mean value of the index of clustering results in K-mediods \& DTW. The $Mean_{CICD}$ is the mean value of the index of clustering results in CICD method. The index value $Max_{K-medoids}$ is obtained when the CICD achieve the max improvement rate in K-mediods \& DTW method. The $Max_{CICD}$ is obtained when the CICD achieve the max improvement rate in CICD method.
\end{tablenotes}
\end{threeparttable}
\end{table*}
With the metrics of $DB$, $S\_Dbw$, $COP$, $VCN$ and $SF$, we evaluated the CICD quality. The smaller $DB$, $S\_Dbw$ and $COP$ are, the higher $VCN$ and $SF$ are, the higher cluster quality is. The Figure~\ref{Fig2}, ~\ref{Fig3}, ~\ref{Fig4}, ~\ref{Fig5}, and ~\ref{Fig6} display the $DB$, $S\_Dbw$, $VCN$, $SF$, $COP$ versus $\gamma$ when our method is used to cluster the LCs. For the comparison,  the results from K-medoids\&DTW is also shown in the Figure~\ref{Fig2}, ~\ref{Fig3}, ~\ref{Fig4}, ~\ref{Fig5}, and ~\ref{Fig6}  after the cluster number of K-medoids\&DTW is set the same as the one of our method that is obtained after the $\gamma$ is set. In other words, the same $\gamma$ means the same cluster number in the Figure~\ref{Fig2}, ~\ref{Fig3}, ~\ref{Fig4}, ~\ref{Fig5}, and ~\ref{Fig6}. When the cluster number are same, the results of CICD has the smaller $DB$, $S\_Dbw$ and $COP$, and the higher $VCN$ and $SF$, outperforming the K-mediods \& DTW in the Figure~\ref{Fig2}, ~\ref{Fig3}, ~\ref{Fig4}, ~\ref{Fig5}, and ~\ref{Fig6}.

Furthermore, the CICD evidently has the significant improvement over the K-mediods \& DTW in $DB$, $VCN$, $S\_Dbw$ and $COP$.  As shown in the Table~\ref{Tab1}, Comparing with the K-mediods \& DTW method the CICD has significant improvement in all cluster validity indices. The CICD has the best improvement in the $VCN$, the mean of the $VCN$ improvement is 184.13\%, and the max improvement is 476.78\%. The CICD has the lowest improvement in the $DB$, but the mean of the $DB$ still has 36.45\% improvement, and the max improvement is 53.82\%. The $DB$, $VCN$, $S\_Dbw$, and $SF$ shown that the max improvement is obtain when the cluster number is small and the $\gamma$ is lager. In addition, the mean of $entropy$ of consumers improvement is 60.87\%, and the max improvement is 68.87\%.

\begin{figure}
\centering
\includegraphics[width=2.45in, height=1.4in]{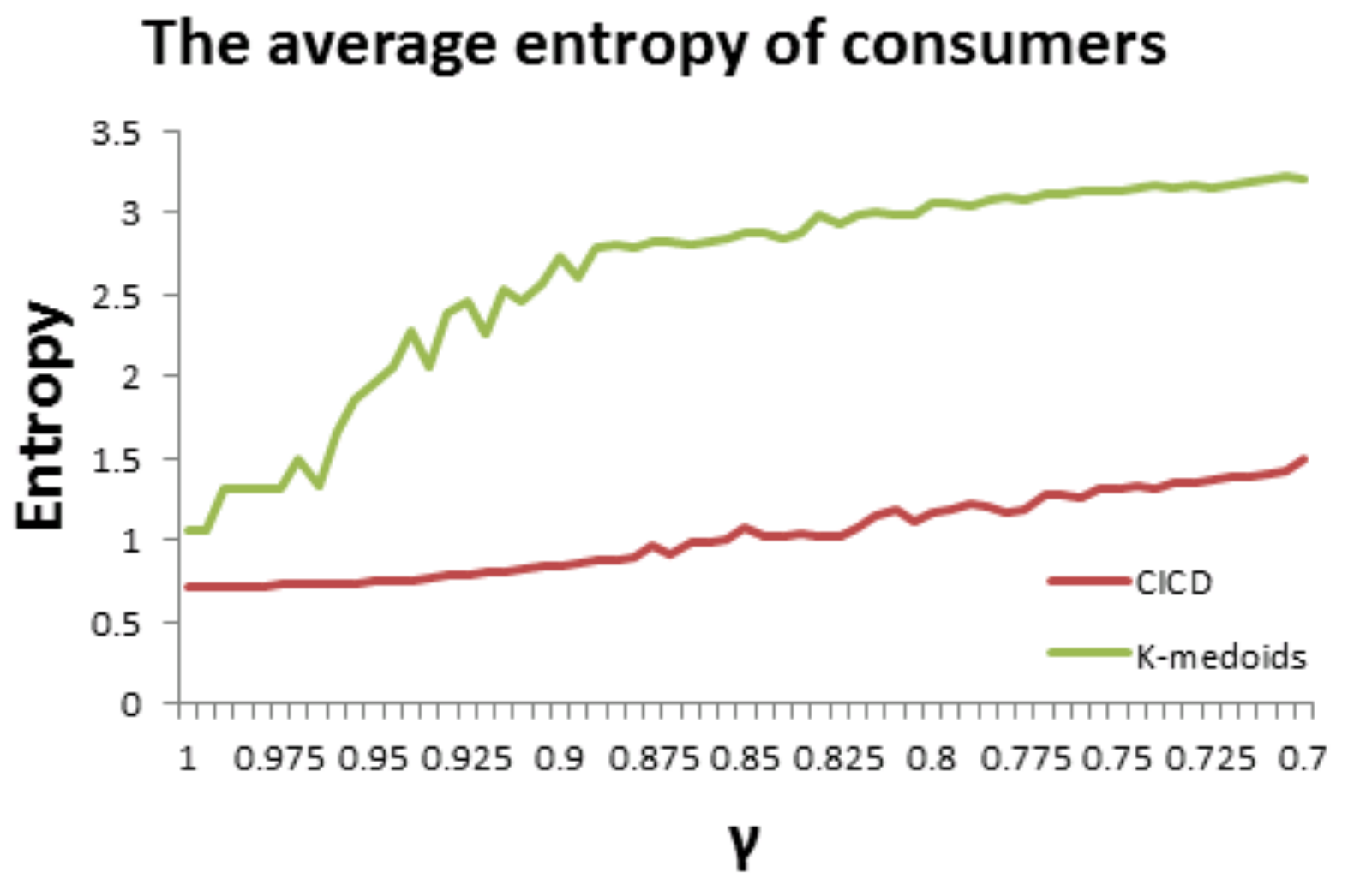}
\caption{\textbf{The average entropy of consumers.} The result of CICD has lower entropy than the result of K-medoids.~\label{Fig7}}
\end{figure}

It is concluded that the lager $\gamma$ is, the more prominent the improvement of the CICD in $DB$, $VCN$, $S\_Dbw$, and $SF$ are, compared with the K-mediods \& DTW. The conclusion demonstrates that the CICD tend to cluster the LCs into less clusters. What's more, the CICD have more lower average $entropy$ of consumers in Figure~\ref{Fig7}, which indicates that the CICD classifies the consumer into more stable representative groups according to the study of the~\cite{teeraratkul2017shape}, which is important to predict individual energy consumption patterns and identify the potential options for the demand respond.

\section{The TLP Directory Construction}

    Ideally, the LCs should be assigned into less cluster, and the LCs in the same cluster are sufficient similar (that mean the variances within a cluster is small), which provide the better understand of the electricity consumption pattern. Actually, it is inevitable that the variances within cluster increase as the number of the cluster decreases, when we applied the clustering technology to cluster the LCs, due to the volatile and uncertain of the electricity consumption behaviors data~\cite{37Li2016Multi}. In practice, the researcher usually make a tradeoff between cluster number and variances within a cluster according to the application requirement.

     In this paper, we proposed a best cluster number determined approach to construct a multi-layers TLP directory.
     In each layer, the variance within the cluster is small when the number of the the TLPs is large, and the variance within the cluster is large when the number of the the TLPs is small.
     In all the layers,  the variance within the cluster decrease when the number of the the TLPs increases. Thus, the researcher is allowed to assess LCs and the customer using an layer or the whole TLP directory according to the application requirement. In detail, we firstly change the $\gamma$ of the Equation~\ref{eq6} to obtain different clustering results in CICD method and extract the TPLs. Secondly, for each clustering result, we calculate its cluster number and the $VCN$ which has been proved that has a excellent performance in determining the optimal number of clusters~\cite{zhou2018novel}. Thirdly, we segment the cluster number into $n$ intervals. For each interval $[i, j)$, we select the TLPs, which correspond to max $VCN$, as a layer of the TLP directory.

\begin{figure}
\centering
\includegraphics[width=3.5in, height=2in]{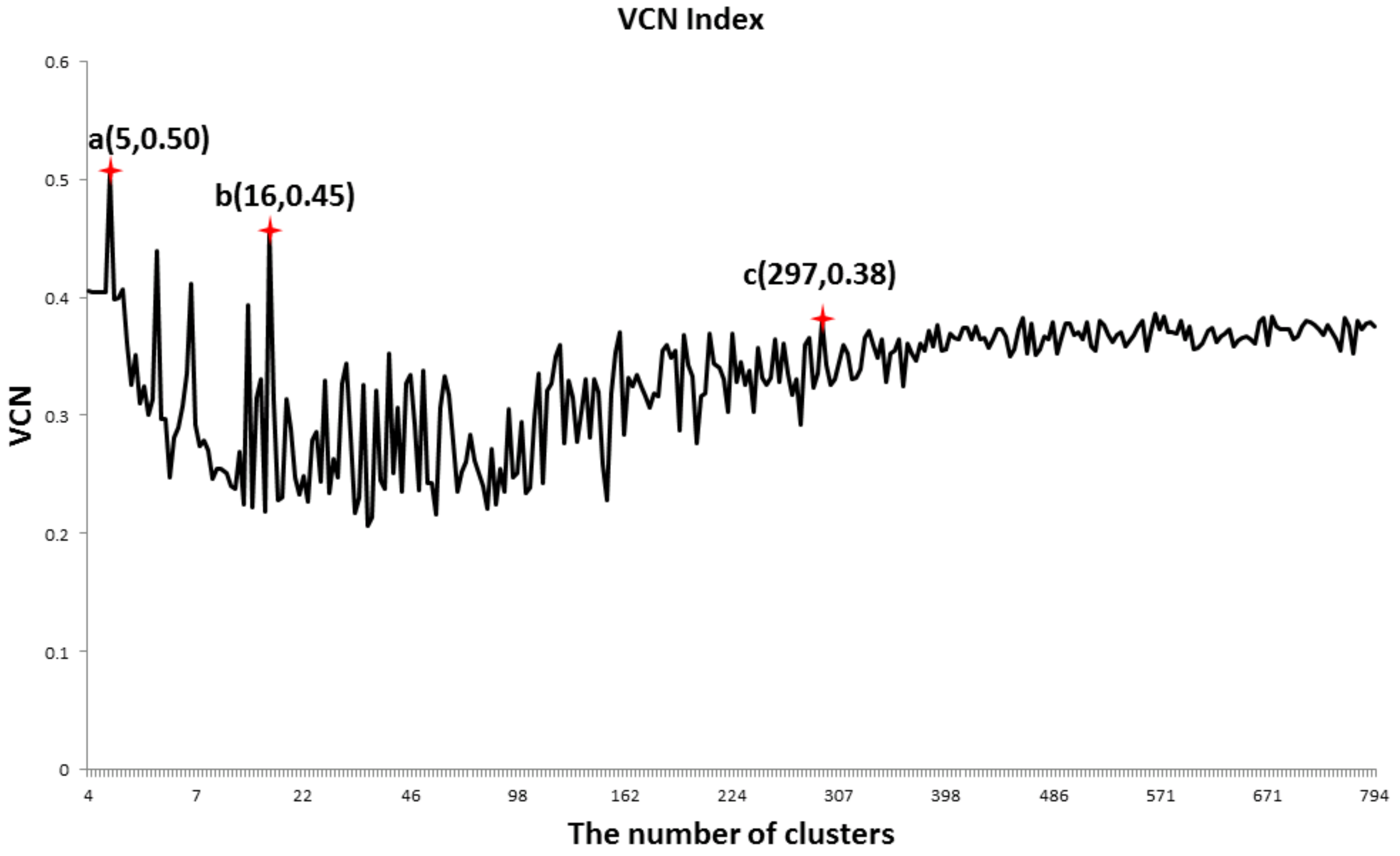}
\caption{\textbf{Selecting the best number of the clusters.} In the intervals $[1,10)$, $[10,100)$ and $[100,\infty)$ we obtain 3 points $a$, $b$ and $c$. Each of the point represents the best cluster number in it's interval.~\label{Fig8}}
\end{figure}

     In this work, we constructed a TLP directory with 3 layers. In order to select the number of clusters for every layer, we change the $\gamma$ of the Equation~\ref{eq6} from $1.00$ to $0.700$ with the interval $0.01$, and calculate the $VCN$ of the clustering result.  As shown in the Figure~\ref{Fig8}, we obtain 3 max value, and each of which corresponds to the best number of the CICD in it's interval. For the first layer, the cluster number is restricted in $[100,\infty)$, and the best number of clusters 297 is obtained when $\gamma=0.832$, the mean variance of which is $0.19$. For the second layer, the cluster number is restricted in $[10,100)$, the best number of clusters $16$ is obtained when $\gamma=0.955$, the mean variance of which is $3.72$. For the third layer, the cluster number is set in $[1,10)$, the best number of clusters $5$ is obtained when $\gamma=0.996$,  the mean variance of which is $11.93$. Thus, a TLP directory with $3$ layers is obtained, the number of the TLPs is respectively $297$, $16$ and $5$ in different layer. And each layer have different level of mean variances within cluster respectively.
     It is observed that the first layer has the maximum cluster number $297$, but the layer has the minimum mean variances within cluster $0.19$ which mean that the LCs within a cluster are most similar to each other and the TLP of the cluster is most similar to the LCs within the cluster\footnote{The TLP is extracting from the LCs with a cluster to represent the cluster. Apparently, when the variance of cluster is small, the TLP is similar to all of the LCs within the cluster.}. Conversely, the last layer has the minimum cluster number $5$, but the layer has the maximum mean variances within cluster $11.93$ which mean that the LCs within a cluster are most differential to each other and the TLP of the cluster is most differential to the LCs within the cluster. The researcher is allowed to assess LCs and the customer using an layer or the whole TLP directory according to the application requirement.

     %Thus, the researcher is allowed to assess LCs and the customer using an layer or the whole TLP directory according to the application requirement.

\section{Conclusion}

   This paper proposed an integrated approach to perform load curves (LC)  clustering.
    First, we proposed a clustering  approach incorporated with  community detection to improve the performance of LC clustering, which includes network construction, community detection and typical load profile  extraction.
    Second, we construct a multi-layer typical load profile (TLP) directory to make the trade-off between variances within a cluster and the number of the clusters.
    In terms of the metrics of five cluster validity indices(Davies-Bouldin index, VCN index, S\_Dbw Index, Score function, and COP index), our method is validated to be effective, outperforming the state-of-the-art methods~\cite{teeraratkul2017shape}.
    %What's more, the TLP directory construction method construct a multi-layers TLP directory, enabling research to assesses the customers combined with the specific application requirements.

    The smart grid network paradigm relies on the exploitation of smart meter data to improve customer experience, utility operations, and advance power management~\cite{quilumba2015using}. Our future work will focus on how to react to demand-response  by analyzing the TLP and the customers. In addition, the external factors (such as customer activity information) will be considered to obtain the features of the customers behaviors behind the LCs.
\section{Acknowledgements}
This work is supported by the Major Program of National Natural Science Foundation of China (Grant No. 61432006), National Key Research and Development Program of China (2016YFB1000600, 2016YFB1000601).

%%%%%%%%reference example%%%%%%%%%%%%%%
%\iffalse

%\begin{thebibliography}{10}

%\end{thebibliography}

%\fi
%%%%%%%%add the reference through ref.bib%%%%%%%%%%%%%%
%\bibliographystyle{ieeetr}
%\bibliography{my}

\end{document}